\documentclass[aps,prl,reprint,groupedaddress,floatfix]{revtex4-1}
\usepackage[pdftex]{graphicx} 
\usepackage{epstopdf}
\usepackage{amssymb} 
\usepackage{amsmath} 

\let\baraccent=\= 
\renewcommand{\=}[1]{\stackrel{#1}{=}} 

\begin{document}
  
\title{Strongly bound excitons dominate electronic relaxation in resonantly excited twisted bilayer graphene}
\author{Hiral Patel$^1$,  Lola Brown$^{2,3}$, Yufeng Liang$^{4}$, Li Yang$^{4}$, Jiwoong Park$^{2,3}$ and Matt W. Graham$^{1}$}
\affiliation{$^{1.}$Department of Physics, Oregon State University, Corvallis, OR USA}
\affiliation{$^{2.}$Department of Chemistry \& Chemical Biology, Cornell University, Ithaca, NY USA}
\affiliation{$^{3.}$Kavli Institute at Cornell for Nanoscale Science, Ithaca, NY USA}
\affiliation{$^{4.}$Department of Physics, Washington University in St. Louis, St. Louis, MO USA}
\begin{abstract}
When two sheets of graphene stack in a twisted bilayer graphene ($t$BLG) configuration, the resulting constrained overlap between interplanar 2$p$ orbitals produce angle-tunable electronic absorption resonances. Using a novel combination of multiphoton transient absorption (TA) microscopy and TEM, we resolve the resonant electronic structure, and ensuing electronic relaxation inside single $t$BLG domains.  Strikingly, we find that the transient electronic population in resonantly excited $t$BLG domains is enhanced many fold, forming a major electronic relaxation bottleneck. 2-photon TA microscopy shows this bottleneck effect originates from a strongly bound, dark exciton state lying $\sim$0.37 eV below the 1-photon absorption resonance.  This stable coexistence of strongly bound excitons alongside free-electron continuum states has not been previously observed in a metallic, 2D material.
\end{abstract}
\maketitle
Photoexcited electrons in graphene relax energetically far faster than the \textit{e-h} separation timescale, making many electronic and optoelectronic applications prohibitive.\cite{malard2013, Graham2013, ju2015, tielrooij2015}  While similar fast, picosecond relaxation timescales are also observed in Bernal stacked bilayer graphene($b$BLG),\cite{Newson2009} slower relaxation might be possible in twisted bilayer graphene ($t$BLG).  In $t$BLG, an off-axis interlayer twist angle ($\theta$) gives rise to band anticrossings and van Hove singularities ($v$Hs, Fig 1b).\cite{Li2009,Lui2010,park2015}  Near such $v$Hs, previous studies show that optical absorption increases by $\sim$20\% and is peaked at an energy, $E_{\theta}$.\cite{Li2009,Lui2010,Jorio2014, Bistritzer2011,Luican2011,Gruneis2008, Havener2012, Wang2010a,Brihuega2012,Havener2014}  This absorption resonance peak increases monotonically with $\theta$ (\textit{see supplementary video}).\cite{Havener2013}  To date however, the properties of photoexcited electrons in $v$Hs  remain unexplored beyond the Raman and linear absorption characterization.  Here, we apply space, time, and energy-resolved 1-photon (1-ph) and 2-photon (2-ph) transient absorption (TA) microscopy to both spectrally map the excited state electronic-structure of $t$BLG and image the ensuing electronic dynamics.

\begin{figure}[h]
\includegraphics[height=1.5in]{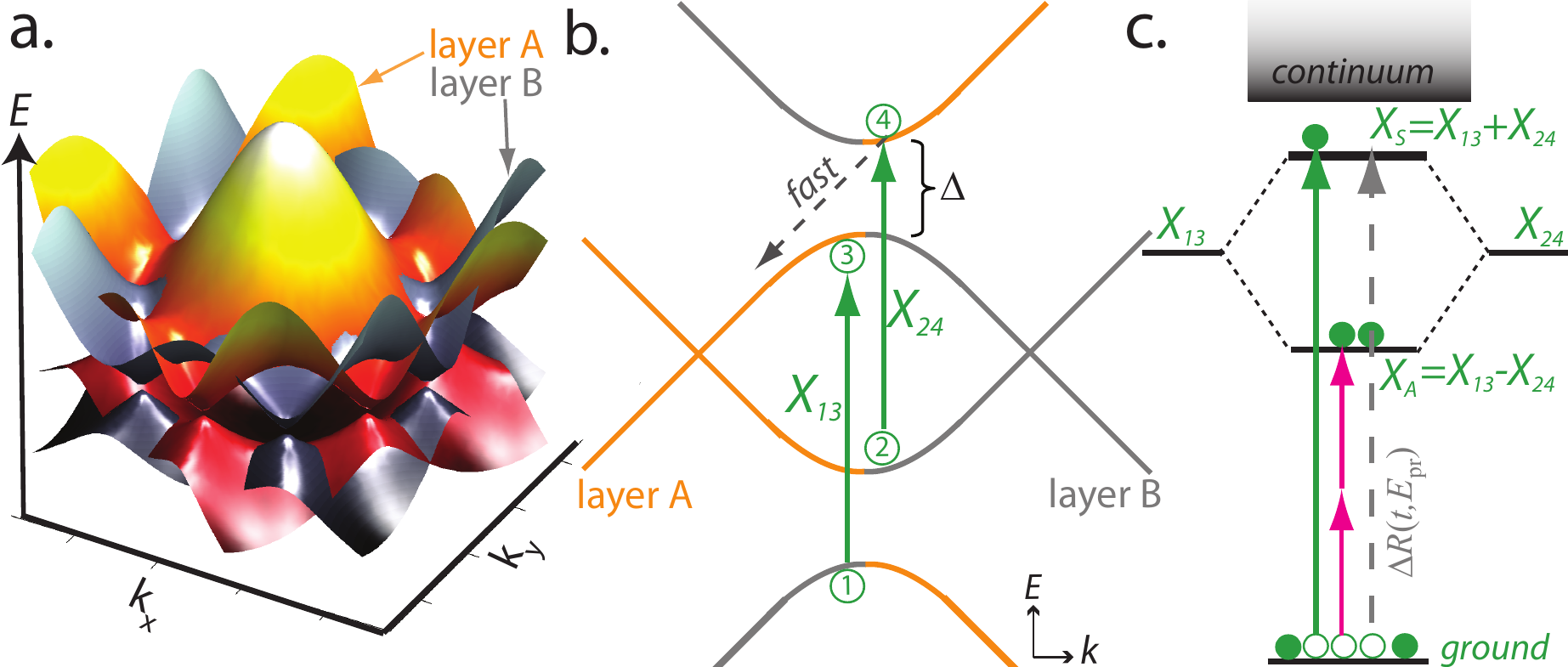}   
\caption{  \textbf{(a)} $t$BLG free-electron interlayer band-structure. \textbf{(b)} Cross-sectional view shows interlayer $v$Hs resonances, $X_{13}$ and $X_{24}$, between band anti-crossing  regions. \textbf{(c)} Alternatively, the degenerate $X_{13}$ and $X_{24}$ states may rehybridize, giving a 1-ph  state above, and a 2-ph allowed exciton state below.  TA (arrows) interrogates the electronic population (circles).
}
 \end{figure}

 

The single-particle band structure for $t$BLG can be understood by superimposing two graphene Brillouin zones, rotated by a twist angle $\theta$, as shown in Fig. 1a.\cite{ Mele2010, LopesdosSantos2007}  The vertical line cutting through the two Dirac points of the graphene layers (Fig. 1b) shows the band anticrossing near the degeneracy with an energy splitting ($\Delta$), and four possible $v$Hs transitions  between the graphene sub-bands labeled 1 through 4.  These optical transitions experience a large joint density of states between the valance bands (1 \& 2) and the conduction bands (3 \& 4), but only 1$\rightarrow$3 (denoted $X_{13}$) and 2$\rightarrow$4 ($X_{24}$) transitions are allowed due to selection rules.\cite{Moon2013, Havener2014}

Outwardly, the $X_{13}$ and $X_{24}$ transitions  shown in Fig. 1b are degenerate $v$Hs similar to  graphene's $M$-point saddle-point exciton.\cite{Mele2010,  Mak2011a} In this case, Coulombic attraction between \textit{e-h} pairs would augment both $X_{13}$ and $X_{ 24}$ transition energies and produce an asymmetric, Fano optical lineshape.\cite{Ohta2012,Fano1961}  Since such unbound Fano excitons couple to continuum states of graphene,\cite{Ohta2012} this model predicts the 1-ph TA response from $t$BLG will decay quickly, with an amplitude and rate similar to single-layer graphene (dotted arrow in Fig. 1b).


Alternatively, previous studies suggest that inclusion of  bound-exciton effects are necessary to simulate the nearly gaussian $t$BLG optical absorption lineshape.\cite{Havener2013, Liang2014b} While one can consider unbound excitonic states for $X_{13}$ and $X_{24}$ independently, such a picture is incomplete because the two states occur at the same energy and momentum, a direct result of the electron-hole symmetry at the $v$Hs in $t$BLG.  A more complete description was given in recent work reported by Liang $et$ $al.$ that predicts formation of stable, strongly bound ($E_B\sim$0.5 eV) excitons.\cite{Liang2014b} These first-principles calculations suggest that interlayer excited states in $t$BLG are better described by renormalized symmetric ($X_S$=$X_{13}$+$X_{24}$) and anti-symmetric ($X_A $=$X_{13}$-$X_{24}$) excitonic states.\cite{Liang2014b} In this model, illustrated in Fig. 1c, $X_S$ corresponds to the optical $t$BLG resonance at $E_{\theta}$, and is an unstable exciton.\cite{Lui2010,Havener2013} Conversely, $X_A$ is only 2-ph accessible and is calculated to be a strongly-bound, localized excitonic state.\cite{Liang2014b}   The remarkable stability predicted for the $X_A$ state results from the deconstructive coherence between the two degenerate Fano resonances rigorously canceling coupling with graphene continuum states. Such a state is termed a 'ghost Fano' resonance. While similar phenomena have been weakly observed in quantum dot and carbon nanotube systems, such strongly bound exciton states have never been observed in a 2D metallic system.\cite{Lu2005, Guevara2003,Liang2014b}  If such ghost Fano excitons are present in $t$BLG, weak exciton-continuum coupling is expected to enhance the local electronic population, giving longer relaxation dynamics for the 1-ph ($X_S$) and 2-ph ($X_A$) TA response.

In this work, we obtain the TA spectra and dynamics of single $t$BLG domains, and map out the different  1-ph and 2-ph electronic transitions predicted by the contrasting $v$Hs and strongly-bound exciton models in Fig. 1.  We further correlate ultrafast TA microscopy with the precise local atomic stacking and grain boundaries, by employing darkfield TEM to definitively assign a twist angle to the absorption resonance, $E_{\theta}$.\cite{Brown2012}   Our experimental TA microscopy approach is outlined in Fig. 2a. The 1-ph TA map in Fig. 2a shows a prominent patch of $6.8^o$ oriented $t$BLG that is surrounded by non-twisted CVD graphene on a silicon nitride membrane substrate. This map was obtained by raster scanning a diffraction-limited pump and probe pulse pair over the graphene. We tuned our 140 fs pump pulse to be resonant with the $6.8^o$ domain at $E_{pump}$$\sim$$E_{\theta}$$\sim$1.3 eV. After a delay time $t$, we detect the differential TA ($\Delta R$($t$) $\propto  \Delta \sigma(t)$) of a collinear probe pulse and construct time-dependent TA maps point-wise. Using probe energies ($E_{pr}$=0.8 eV) well below the resonance $E_{\theta}$, in Fig. 2a  graphene gave an interband decreased absorption response (i.e. Pauli blocking of probe beam) everywhere at all time delays.\cite{Graham2013}   

  \begin{figure}[htbp]
\includegraphics[height=4.8in]{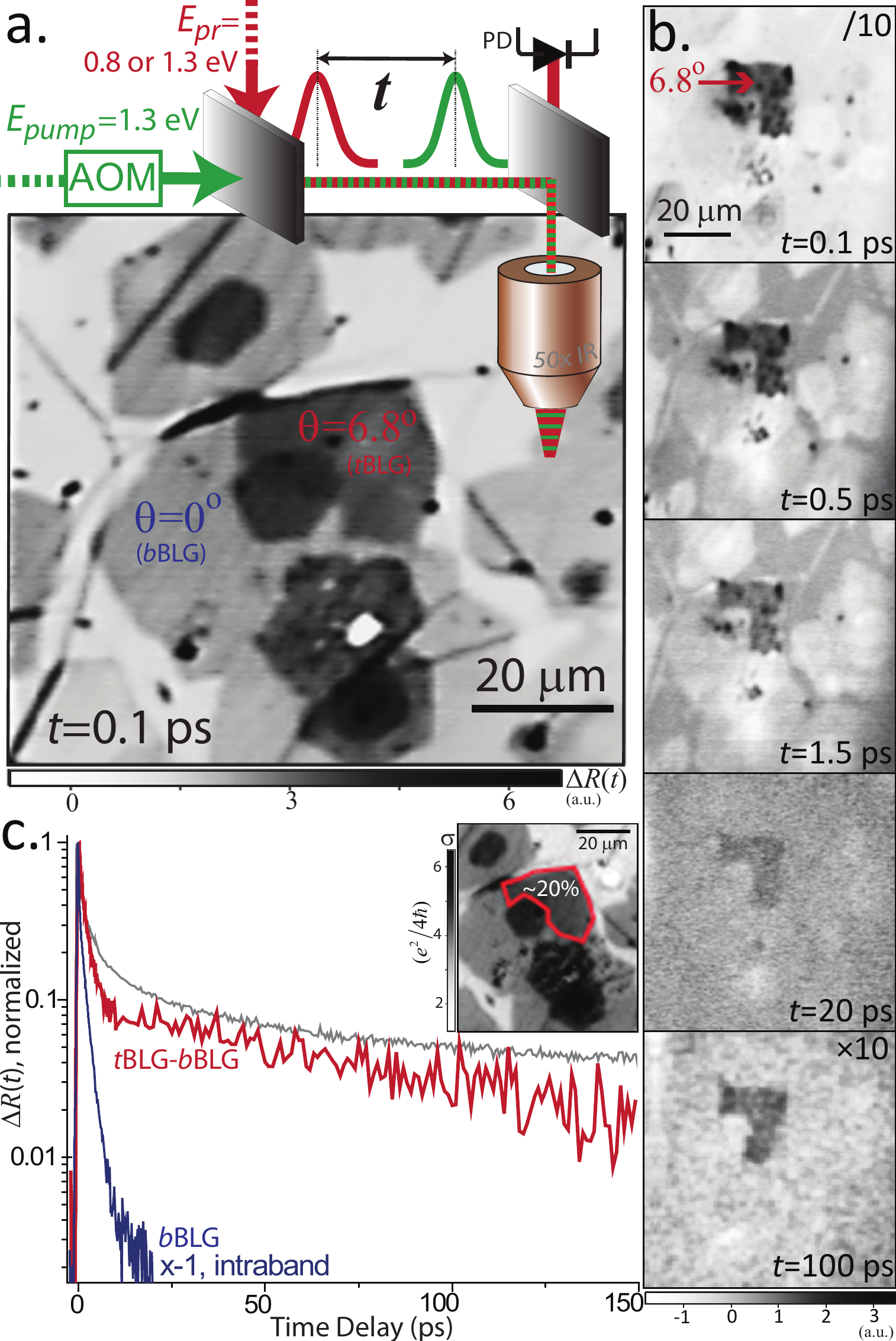}   
\caption{ \textbf{$t$BLG electronic relaxation bottleneck}   \textbf{(a)}  Ultrafast scanning TA microscopy map of  multilayer graphene.  The TA Pauli blocking response is $>$2$\times$ enhanced for the $6.8^o$ $t$BLG domain.  \textbf{(b)}  On-resonance TA maps at $E_{pump} \sim E_{pr}=E_{\theta}$ show a TA response localized to the $6.8^o$ domain is present for $t$$>$100 ps . The surrounding graphene gives only a weak transient intraband response (\textit{opposite sign}). \textbf{(c)}  TA relaxation kinetics of the $b$BLG  and $t$BLG regions labeled (vs. SWNT $E_{11}$ state, \textit{gray}).  Corresponding linear absorption map at 1.3 eV shows $t$BLG (\textit{red}) is only $\sim$20\% stronger than in $b$BLG \textbf{(\textit{inset})}.} 
\end{figure}

Our TA maps can be interpreted as 'movie frames' that closely approximate the relative photoexcited electronic population at a particular probe energy and time-delay (\textit{see supplementary video 2}).  The 6.8$^o$ $t$BLG region labeled in Fig. 2a has a $\sim$two-fold stronger TA Pauli blocking response than the adjacent $0^o$ stacked regions. However, the corresponding linear absorption map in Fig. 2c only shows a $\sim$20\% resonant enhancement.  To account for this discrepancy, electrons in interlayer $t$BLG avoided crossing regions must relax much slower than the surrounding non-twisted graphene bilayers, suggesting an intrinsic electronic relaxation bottleneck.  

\begin{figure*}[htbp]
\includegraphics[height=4in]{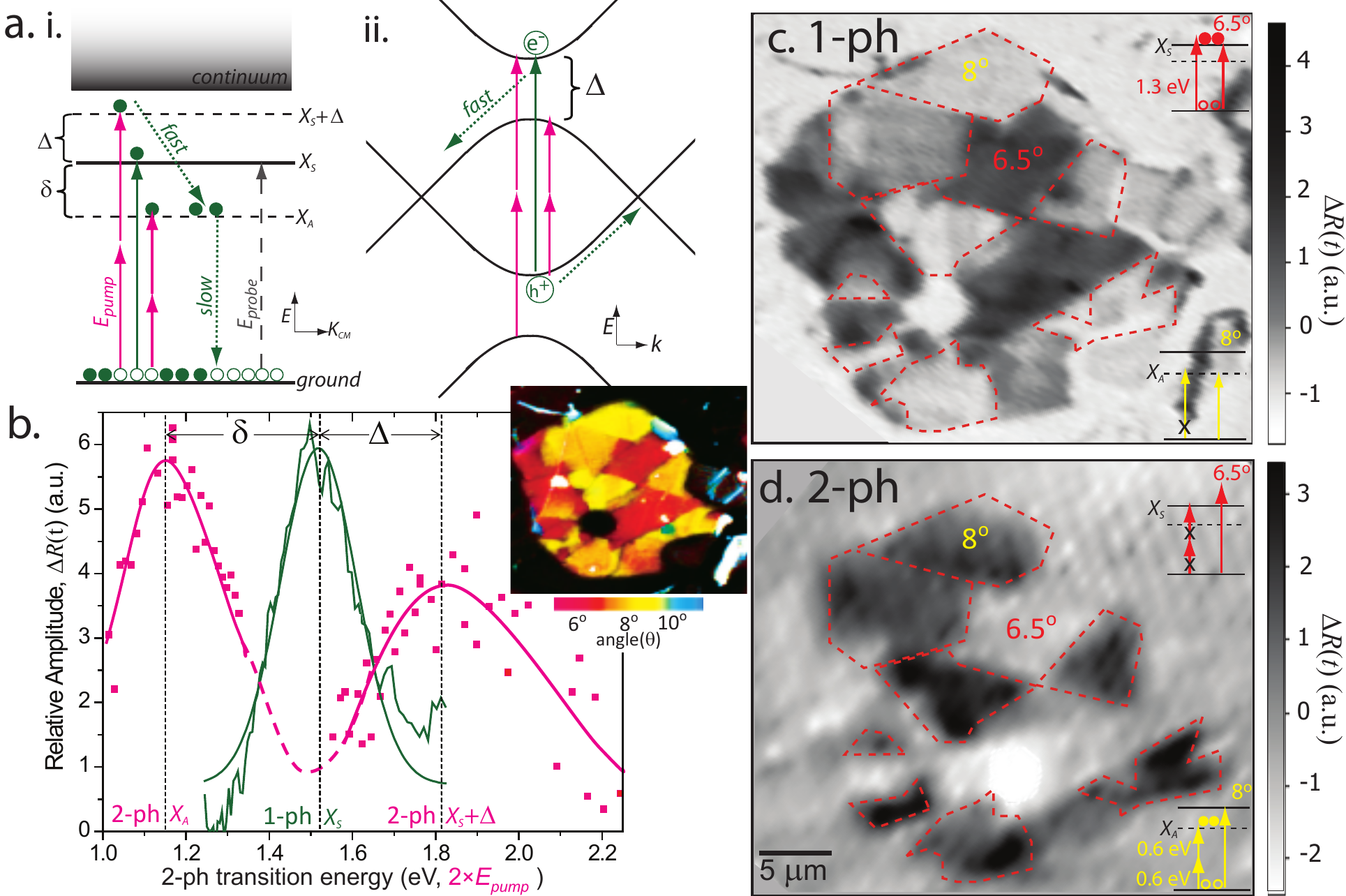}   
\caption{\textbf{1 photon vs. 2 photon absorption}  of  $\sim$$6.5^o$ and $\sim$$8^o$(dotted outlines) $t$BLG domains. \textbf{(a)}  Competing models; (\textit{i.})bound exciton model, (\textit{ii.})continuum model  \textbf{(b)} TA spectrum (\textit{red})  vs.  2-ph transition pump energy.   1-ph interlayer absorption spectrum ($\sigma_{tBLG}$-$\sigma_{bBLG}$,  \textit{green}) for 8$^o$ $t$BLG.   The lowest 2-ph peak fits best to a gaussian lineshape centered at $\delta=$0.37 eV below $E_{\theta}$.  There is also a Fano resonance lineshape feature at $\Delta=$ 0.33 eV above. (\textit{inset}, map of $t$BLG absorption resonance vs. twist angle)  \textbf{(c)} 1-photon TA map, at $E_{pump}$=1.3 eV shows a strong electronic bleach only from the resonantly excited $6.5^o$ domains.  \textbf{(d)} Conversely, a 2-ph TA map  at $E_{pump}$=0.6 eV shows a ground-state bleach only from the $8^o$ domains.  Combined, these maps demonstrate that the ($X_{S}$) state is 1-ph allowed and ($X_{A}$) state is only 2-ph allowed.   
   }
\end{figure*}
 
In Figure 2b, we repeat the 1-ph measurement but instead \textit{resonantly probe} the electronic population at $E_{\theta}$  ($E_{pump}$=1.33  eV, $E_{pr}$=1.26 eV probe). Compared against the corresponding linear absorption map in Fig. 2c (\textsl{inset}), the TA maps differ in both sign and absolute amplitude.   Strikingly, only the $6.8^o$ $t$BLG domain gives a strong TA Pauli blocking response.  Meanwhile, the surrounding graphene in Fig. 2b  gives a weak, short-lived graphene intraband TA response signified by its \textit{opposite sign}.  This suggests that interlayer $t$BLG electrons are decoupled from the intraband transient response that dominates the TA map everywhere else in Fig. 2b.  Surprisingly, the subsequent TA movie frames show excited carriers are present even $>$100 ps after initial excitation.  Both observations definitively show that interlayer electrons excited at $E_{\theta}$ experience a major bottleneck restricting electronic relaxation.  Such a strong and long-lived electronic signal in $t$BLG disagrees with the continuum Fano resonance model (Fig. 1b), but can be explained by an excitonic model (Fig. 1c) where weak exciton-continuum coupling allows for stable exciton formation.\cite{Liang2014b}

To isolate the relaxation rates intrinsic to interlayer $t$BLG electrons excited at $E_{\theta}$, we plot $t$BLG-$b$BLG (\textsl{red}) in Fig. 2c  by subtracting the much weaker (and opposite signed) intralayer electronic TA response (\textsl{blue}).  A similar approach has been previously used to decouple linear absorption spectra, as $\sigma_{tBLG}$-$\sigma_{bBLG}$.\cite{ Havener2012,Havener2014}   A least-squares deconvolution exponential fit of the kinetic decay requires only a biexponential fit that decays with lifetimes of 1.4$\pm$0.1 ps and 66$\pm$4 ps. These lifetime components are remarkably long for any electronic state within a metallic system.  By repeating our linear and TA measurements at low temperatures,  we further found the interlayer electronic response appears largely invariant to both lattice temperature (5-295 K) and the substrate used (\textit{see supplemental materials}). Furthermore, the TA response did not shift sign as the probe wavelength was scanned through the resonance, $E_{\theta}$.  Combined, these observations suggest that laser-induced heating effects do not contribute appreciably to the overall large TA signal response, and so the TA signal is predominately electronic in origin.  

If carrier in $t$BLG are unbound excitons, the TA (\textit{red}) in Fig. 2c  must relax at a rate similar to $b$BLG (\textsl{navy}), providing that phonons with $E$$>$$\Delta$ (dotted arrow in Fig. 1b) are available to scatter carriers through the anti-crossing gap ($\Delta$) illustrated in Fig. 1b.\cite{Graham2013} Comparison of the short-time $t$BLG kinetics against $b$BLG in Fig. 2c (\textit{red}) reveals the absence of the dominant fast sub-ps electron relaxation components associated with graphene electron thermalization and optic phonon emission.\cite{Wang2010a}  Remarkably, the shortest interlayer $t$BLG lifetime is 1.4 ps, which is similar to graphene's rate-limiting relaxation rate that is often associated with disorder-assisted or supercollision relaxation.\cite{Graham2013,Graham2013a}  The absence of the sub-ps relaxation processes, and the emergence of this long $\sim$66 ps decay in $t$BLG relaxation kinetics, suggests that some electrons are  decoupled from graphene's continuum states, as predicted by the strongly bound exciton model.\cite{Liang2014b}    

The unexpected TA bottleneck we observe in $t$BLG may be compared against semiconducting single-walled carbon nanotubes (SWCNTs), a similar carbon system with constrained 2$p$ orbital interactions. It is established that SWCNTs have 1-ph and 2-ph excitonic states resulting from (chiral) angle dependent overlapping 2$p$ orbitals.\cite{Srivastava2008,Matsunaga2008, Deslippe2007}  Fig. 2c (\textsl{gray}) directly compares the $E_{11}$ exciton relaxation rate of (6,5) chirality SWCNTs against $t$BLG  (\textsl{red}).  While the short time behavior differs greatly, Fig. 2c shows the longer components of both traces decay at a similar rate, suggesting that the dynamic phonon environment causing $E_{11}$ exciton relaxation in SWCNTs might be of a similar nature to the interlayer exciton-phonon interactions causing exciton relaxation in $t$BLG.   



We can better distinguish between competing $v$Hs  and bound exciton models outlined in Fig. 3a, by exploiting the  2-ph selection rules required for the predicted dark $t$BLG exciton state, $X_A$.\cite{Moon2013,Liang2014b}  To search for possible dark state transitions, we used a different sample of CVD bilayer graphene.  The linear absorption map shown in Fig. 3b (\textit{inset}) reveals a series of $t$BLG domains with twist angles of either $\sim$$6.5^o$ (\textit{yellow}) or $\sim$$8^o$ (\textit{red}), corresponding to $E_{\theta} \sim$1.25 eV and 1.55 eV respectively.  Figure 3b plots the 2-ph TA spectrum obtained using IR pump energies ranging from $E_{pump}=$0.49 to 1.15 eV, and a $8^o$ resonant probe at $E_{\theta}$$\sim$$E_{pr}$=0.56 eV.   We observe two clear TA peaks centered at 1.18 eV and 1.82 eV that originate from a resonantly enhanced 2-ph absorption. Specifically, as illustrated in Fig. 3ai,  we observe these dark states through resonant 2-ph enhanced Pauli blocking of the depleted ground state.  Moreover, our ability to probe electronic population of optically dark state requires that $X_S$ and  $X_A$ states share a common ground state; an inherent feature of a bound exciton model.\cite{Deslippe2007}   

The lowest peak in Fig. 3b indicates that enhanced 2-ph absorption took place via a discrete, low-lying transition centered at 1.18 eV.   Comparing the 2-ph peak against the 1-ph absorption resonance at $E_{\theta}$=1.55 eV  (green, Fig. 3b), we readily obtain the energy-state splitting parameters of $\delta=0.37$ eV and $\Delta=0.33$ eV.  This 0.37 eV energy splitting closely matches the theoretically predicted $\delta \cong$0.4-0.5 eV, state splitting calculated for 21$^o$ $t$BLG.\cite{Liang2014b}    Such a large bright-dark state energy splitting is much greater than the analogous state splitting in SWCNTs,  and explains why  photoluminescence has not yet been observed from resonantly excited $t$BLG domains.  
 
To completely map the selection rules associated with $t$BLG electronic transitions, we compare the 1- and 2-ph TA microscopy response of $6.5^o$ and $8^o$ oriented domains in Fig. 3c-d. Fig. 3c maps-out the 1-ph TA response for $E_{pump}$=1.3 eV, $E_{pr}$=1.2 eV at $t$=$0.3$ ps.    Despite the $X_A$ state being 1-ph resonant with the $\sim$$8^o$ $t$BLG (dotted red outlines), we observed only a weak intraband response as was seen for $b$BLG regions previously (Fig. 2b).  This confirms that the $X_A$ transition is not 1-ph accessible. In contrast, the 6.5$^o$ $t$BLG domains give a strong Pauli blocking response because $E_{pump}$ is resonant with $X_S$.  
 
%

 2-ph resonant transitions of single $t$BLG domains are imaged in Fig. 3c, by tuning our pump pulse energy to roughly half the predicted $8^o$ $X_A$ state energy(see Fig. 3a), or $E_{pump}$= 0.6 eV.  Comparison of the TA maps in Fig. 3c against 3d show all of the 8$^o$ $t$BLG domain excitations that were forbidden under 1-ph excitation conditions are now allowed for a 2-ph excitation. Conversely, all the 6.5$^o$ $t$BLG domain excitations that were observed under 1-ph excitation conditions now appear dark (inaccessible) under two-photon excitation.   Using state parity, we assign the  two-photon accessible dark states in Fig. 3d to electronic carriers populating the $X_A$ state of 8$^o$ $t$BLG.   Together, Fig. 3c and 3d show that the $X_S$ bright state is two-photon forbidden, and the dark $X_A$ state is only two-photon allowed.  These strongly enforced selection rules follow the parity expectations of a roughly hydrogenic-like, strongly-bound exciton model advocated by recent first-principle simulations.\cite{Liang2014b}

The 2-ph spectral peak centered at 1.82 eV in Fig. 3b  has not been previously predicted or observed. This peak has a broader, asymmetric shape, that fits  better to a Fano lineshape expected from the unbound exciton model.\cite{Mak2011,Chae2011} In contrast, the other 1-ph and 2-ph peaks in Fig. 3b fit best to a gaussian lineshape, a common characteristic of bound excitonic transitions.  Accordingly,  we infer that the two-photon absorption near 1.82 eV is best assigned to an unbound state transition labeled $X_{S}$+$\Delta$ in Fig. 3ai.  Conversely, the two-photon absorption resonance lying $\delta=0.37$ eV below is best characterized as the  $X_{A}$ bound exciton or ghost Fano resonance peak predicted by Liang et al. \cite{Liang2014b} as supported by (i.) its asymmetric energy spacing (i.e. $\delta$ vs. $\Delta$), (ii.) gaussian lineshape, (iii.) long electronic lifetime and (iv.) parity enforced two-photon selection rules for the $X_A$ and $X_S$ transitions. 

\begin{figure}[htbp]
\includegraphics[height=3in]{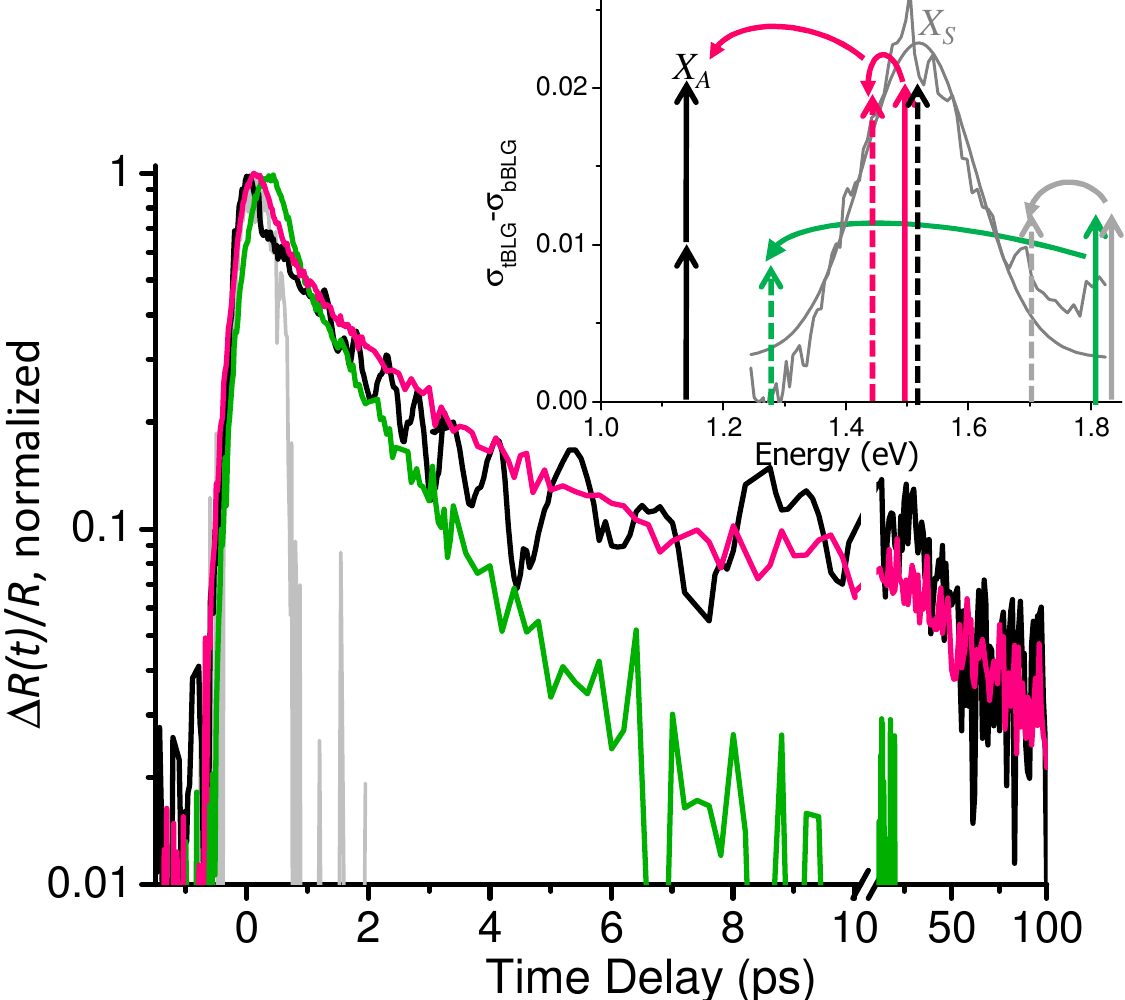}   
\caption{ \textbf{Resonant vs. non-resonant electronic relaxation}  for $8^o$ $t$BLG.   The vertical pump (\textit{solid arrows}) and probe (\textit{dashed arrows}) combinations labeled on the  absorption spectrum (\textbf{\textit{inset}}) of $8^o$ $t$BLG , correspond by colour to the normalized 1- or 2-ph. relaxation kinetics plotted.  
 }
\end{figure}  

Using both explicit calculations based on the Bethe-Salpeter equation and effective low-angle continuum model, Liang \textit{et al.}  predicted radically different electronics properties emerge for both the theorized $X_S$ and $X_A$ exciton states.\cite{Liang2014b} Specifically,  the symmetric $X_S$ state was found to have delocalized wavefunctions and a negligible binding energy. Conversely, the antisymmetric state $X_A$ is predicted to be optically dark, insensitive to $e-h$ charge screening, and strongly bound.\cite{Liang2014b}   While certain phonons can scatter bound excitons into the lower-lying continuum states, the exciton-continuum coupling for the $X_A$ is predicted to be vanishingly small, and roughly intensive to charge screening effects.\cite{Liang2014b}  Accordingly, both theory and our TA microscopy now support that fast exciton dissociation becomes unfavorable in the $X_A$ state of $t$BLG, enabling stable and meta-stable bound exciton states to form.

Lastly, we compare the 1-ph $X_S$ (\textit{pink}) vs. 2-ph $X_A$ (\textit{black}) electron relaxation kinetics measured for 8$^o$ $t$BLG in Fig. 4.  We find that  the normalized TA relaxation kinetics for one-photon and two-photon resonant excitations are nearly identical.  These matching kinetics indicate that both signals originate from the same depleted common ground state, and that the electrons are impulsively transferred from the bright $(X_{S})$ to the dark $(X_{A})$ state as illustrated in Fig. 3ai.  Such fast kinetic $X_{S}$$\longrightarrow$$X_{A}$ relaxation  is consistent with theory showing that $X_S$ is an unstable exciton state.\cite{Liang2014b}  As a control, in Fig. 4 (\textit{gray}) we show that the relaxation kinetics are impulsive when both $E_{pump}$ and $E_{pr}>E_{\theta}$ , indicating that only free electron states are probed above resonance.  Lastly we consider  the case of $E_{pr}<E_{\theta}<E_{pump}$, and find  the long-decay components are consistently absent, suggesting resonant optical excitation may be required to form a long-lived stable exciton.    Nonetheless, the off-resonance $t$BLG kinetic relaxation (\textsl{green}) is still significantly enhanced in amplitude and lifetime compared to the $b$BLG TA response (Fig. 2c).  This suggests that an electronic relaxation bottleneck effect is still present even when $t$BLG is optically excited above resonance.

By definitively isolating the interlayer electronic dynamics of 1- and 2-ph resonant optical transitions in $t$BLG, we have uncovered a fine-structure of bound  ($X_A$) and unbound ($X_S$, $X_S$+$\Delta$) exciton states that agrees well with recent simulations.\cite{Liang2014b} Specifically, we employed a novel form of diffraction-limited TA microscopy to obtain the intrinsic spectra and dynamics of single $t$BLG domains under a variety of resonant and non-resonant pump/probe combinations.  In Fig. 2b, we show a TA-movie of electronic population that reveals the striking contrast between the bound-exciton carriers in the $t$BLG region and the free-electron population in the surrounding graphene.  These results suggest that the photoexcited $t$BLG  interlayer electrons are initially decoupled from scattering into graphene continuum sates, and experience a significant electron relaxation bottleneck.   In particular, resonantly excited carriers in $t$BLG give much stronger TA amplitude with longer relaxation kinetics for both the short ($\sim$2 ps) and long relaxation timescales ($\sim$70 ps). This bottleneck is best explained by the existence of strongly-bound excitons in a 'ghost Fano' state that we explicitly resolve using 2-ph TA microscopy.\cite{Liang2014b}    Our results imply that $t$BLG may be a unique hybrid electronic material where free-electron metallic character can coexists alongside stable exciton states. The work further opens up possible new avenues for carrier extraction that combine the high conductivity of metallic \textit{intra}layer-electrons, with the enhanced electronic population that is now established for the \textsl{inter}layer electrons in $t$BLG.

\textbf{\textit{Methods}}: Multi-layer graphene was grown using low pressure CVD method on copper foil and transferred to silicon nitride grids (\textit{see supplemental materials}).\cite{li2011}  Areas containing low-angle $t$BLG  were first identified using, a combination of hyperspectral absorption imaging technique, and dark field TEM (DF-TEM).\cite{Havener2013}  Final twist angle assignments of the bilayer patches were made by correlating the linear absorption and 1-ph TA peak spectral peaks energies.

$t$BLG bright ($X_{S}$) and dark ($X_{A}$) states and their corresponding electronic dynamics were measured using  1- and 2-ph confocal scanning TA microscopy.\cite{hartland2010}  Collinear pump-and probe pulses were obtained from two independently tunable outputs of an ultrafast system composed of Ti:Saph oscillator (Coherent Chameleon Ultra II, 80 MHz, wavelength range 680-1080 nm) pumping an optical parametric oscillator (APE-Compact, wavelength range 1000-4000 nm).  For one-photon TA measurements requiring pump and probe pulse doubly resonant with the bright ($X_{S}$) transition, a white-light supercontinuum probe was instead used.  Cross-correlation of the pump and probe after the objective yielded a FWHM  pulse duration of $142$ fs.     

After a mechanical delay stage, both the pump, and the probe beams were aligned in a collinear geometry, raster-scanned by piezo-scanning mirror and coupled into a confocal scanning microscope via a 50X IR-region enhanced, achromatic objective (NA= 0.65).  One- and two photon transient absorption signals were detected by measuring the probe beam on with a TE cooled InGaAs detector connected to a Zurich HF2LI lock-in amplifier. The pump beam was modulated at either 0.25 or 1 MHz using a AO-modulator (Gooch \& Housego) to enable high-frequency lock-in detection of the differential reflectivity. Appropriate optical filters were used in front of the detector to block the pump beam. The pump and probe spot sizes on the sample were  determined to $\sim$1.5 $\mu$m, by fitting to a confocal scanning reflection profile of deposited gold pads. The fluence of the probe power was ~5\% of the pump fluence.  Except where specified, all the measurements were done at 295 K. The probe power was  fixed at ($\sim 1 \times 10^{12} $ photons/cm$^2$ ) for the pump power dependence measurements. Microscope objective/transmission corrections curves were measured and rigorously taken into account for all the wavelengths, after each measurement.

\textbf{Acknowledgments}: this research was supported by the Oregon State Foundation and Cornell's AFOSR (FA 9550-10-1-0410) grant. This work made use of TEM facilities of Cornell Center for Materials Research Shared Facilities which are supported through the NSF MRSEC program (DMR-1120296). We gratefully acknowledge Robin Havener. 


%

\end{document}